\documentclass{nature}
\usepackage{graphicx}
\usepackage{physics}
\usepackage{color}
\usepackage[utf8]{inputenc}
\makeatletter
\let\saved@includegraphics\includegraphics
\AtBeginDocument{\let\includegraphics\saved@includegraphics}
\renewenvironment*{figure}{\@float{figure}}{\end@float}
\makeatother

\title{Experimental measurement of the divergent quantum metric of an exceptional point}

\author{Qing Liao$^{1}$\footnote{email:liaoqing@cnu.edu.cn} , Charly Leblanc$^3$, Jiahuan Ren$^{1,2}$\footnote{These authors contributed equally: Q. Liao, C. Leblanc,  J. Ren}, Feng Li$^{4}$\footnote{email:felix831204@mail.xjtu.edu.cn}, Yiming Li $^{4}$,  \\ Dmitry Solnyshkov$^{3,5}$\footnote{email:dmitry.solnyshkov@uca.fr}, Guillaume Malpuech$^{3}$, Jiannian Yao$^{2}$, Hongbing Fu$^{1,2}$\footnote{email:hbfu@cnu.edu.cn}}

\begin{document}
\maketitle

\begin{affiliations}
\item Beijing Key Laboratory for Optical Materials and Photonic Devices, Department of Chemistry, Capital Normal University, Beijing 100048, People's Republic of China
\item Tianjin Key Laboratory of Molecular Optoelectronic Sciences, Department of Chemistry, School of Sciences, Tianjin University, Collaborative Innovation Center of Chemical Science and Engineering, Tianjin 300072, People's Republic of China
\item Institut Pascal, PHOTON-N2, Universit\'e Clermont Auvergne, CNRS, SIGMA Clermont, F-63000 Clermont-Ferrand, France.
\item Key Laboratory for Physical Electronics and Devices of the Ministry of Education \& Shaanxi Key Lab of Information Photonic Technique, School of Electronic Science and Engineering, Faculty of  Electronic and Information Engineering, Xi’an Jiaotong University, Xi’an 710049, China
\item Institut Universitaire de France (IUF), 75231 Paris, France
\end{affiliations}

\begin{abstract}
The geometry of Hamiltonian's eigenstates is encoded in the quantum geometric tensor (QGT) \cite{provost1980riemannian,berry1989quantum,Berry2020}. It contains both the Berry curvature, central to the description of topological matter and the quantum metric. So far the full QGT has been measured only in Hermitian systems \cite{Yu2019,gianfrate2020measurement}, where the role of the quantum metric is mostly shown to determine corrections to physical effects \cite{peotta2015superfluidity,Liang2017,Gao2014,Piechon2016,PhysRevLett.115.166802}. 
On the contrary, in non-Hermitian systems, and in particular near exceptional points \cite{Moiseyev2011,Konotop2016,Ganainy2018,Ozdemir2019}, the quantum metric is expected to diverge \cite{Brody2013} and to often play a dominant role, for example on the enhanced sensing \cite{chen2017exceptional} and on wave packet dynamics \cite{Solnyshkov2020arxiv}. In this work, we report the first experimental measurement of the quantum metric in a non-Hermitian system. The specific platform under study is an organic microcavity with exciton-polariton eigenstates, which demonstrate exceptional points. We measure the quantum metric's divergence and we determine the scaling exponent $n=-1.01\pm0.08$, which is in agreement with theoretical predictions for the second-order exceptional points.    
\end{abstract}

\maketitle


The recent development of experimental techniques and theoretical understanding has allowed to measure both components of the quantum geometric tensor (the Berry curvature and the quantum metric) experimentally \cite{Yu2019,gianfrate2020measurement}. In particular, the use of optical systems allows to access the non-trivial geometry of real photonic bands and to observe the related consequences on wave packet propagation and the anomalous Hall effect \cite{gianfrate2020measurement}.

At the same time, the studies of non-Hermitian systems \cite{Moiseyev2011,Konotop2016,Ganainy2018,Ozdemir2019} have also started to deal with the topology of the exceptional points, which are the branch points of the multi-valued Riemann surface formed by the eigenvalues of the Hamiltonian of such systems. It was shown that the chiral dynamics associated with this non-Hermiticity is extremely promising for applications \cite{Milburn2015,Doppler2016}. Crucially, the good topological invariant in vicinity of these points is not anymore associated with the Berry curvature of the eigenstates, but with the winding number of the so-called effective field \cite{Bliokh2017,Shen2018} (and the associated complex eigenvalues), determined by the Hermitian and non-Hermitian parts of the Hamiltonian itself. Indeed, because of the non-Hermitian contribution, the adiabatic description of dynamics based on the Berry curvature becomes irrelevant \cite{Berry2011a,Berry2011b}. 
On the other hand, the quantum metric should exhibit a hyperbolic divergence at the exceptional points of 2nd order (with square root topology) \cite{Brody2013}. This divergence has remarkable physical consequences for the dynamics of wave packets centered at exceptional points \cite{Solnyshkov2020arxiv}. Here, the quantum metric is not responsible for small corrections, it has a dominant role, determining a non-vanishing constant group velocity with a polarization-dependent direction. However, the quantum metric of a non-Hermitian system has never been measured experimentally so far, in spite of the extended studies of such points in optics \cite{Berry2003,Ozdemir2019} which date back to Voigt \cite{Voigt1902}, and of their recent observation in microcavities \cite{Richter2019}. 

In this work, we study the modes of an organic microcavity \cite{Microcavities} exhibiting a polarization-dependent strong coupling, which provides a pronounced non-Hermitian response ensuring well-defined exceptional points. We measure the Stokes parameters of the eigenmodes in vicinity of the exceptional points and extract the corresponding quantum metric. We demonstrate that this metric is diverging, exhibiting a scaling exponent $n=-1.01\pm0.08$. The coefficients of the measured hyperbola correspond to the analytical predictions based on an effective Hamiltonian.

\begin{figure}[tbp]
\begin{center}
\includegraphics[width=12.5cm]{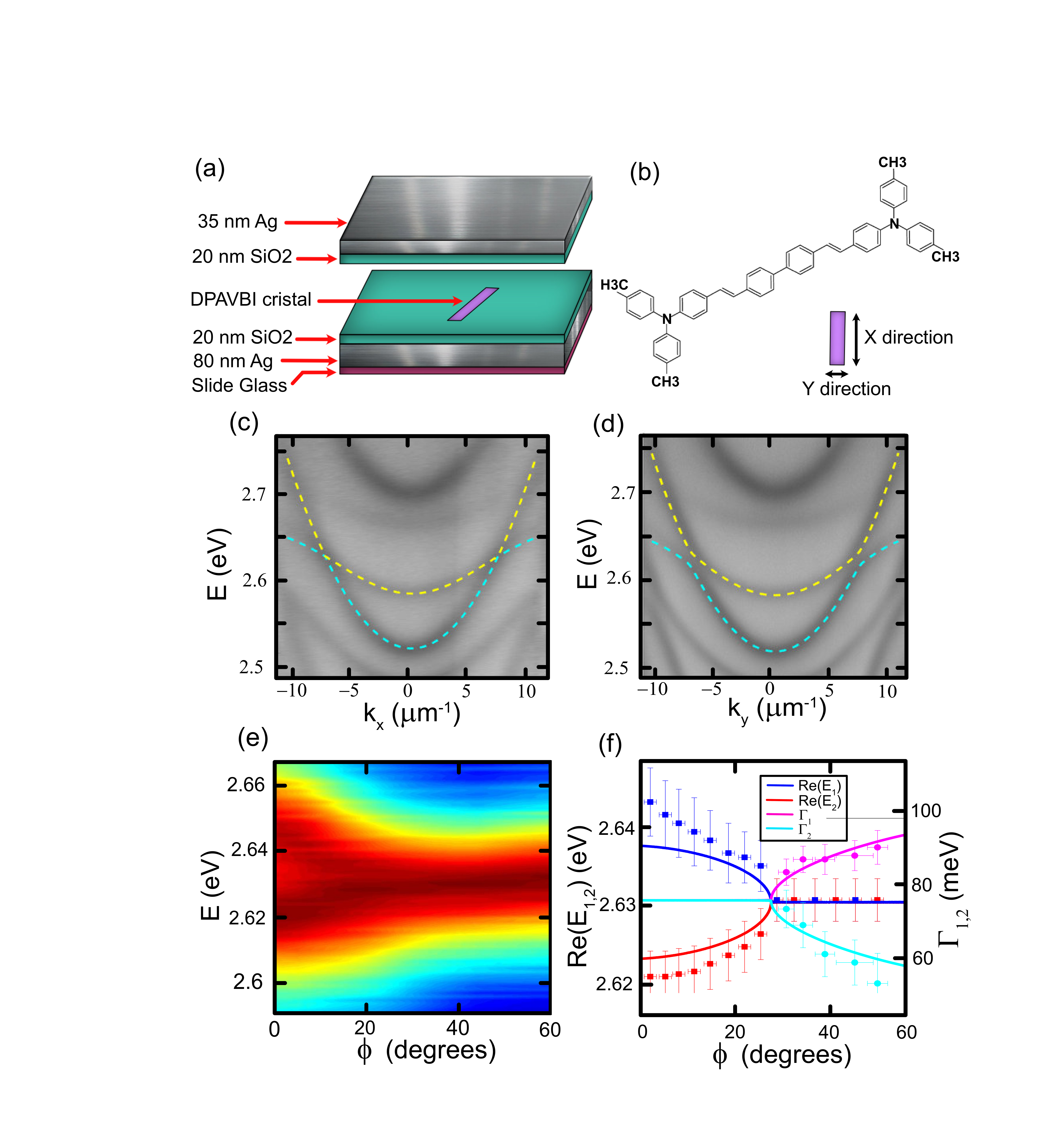}
\caption{\textbf{Reflectivity of the organic microcavity.} a) Scheme of the microcavity sample. b) Structure of the DPAVBi molecule. c), d) Reflectivity as a function of wave vector $k_x$ and $k_y$ (respectively) and energy, exhibiting anticrossing along $k_y$. e) Reflectivity as a function of the in-plane polar angle $\phi$ for $|k|=|k^*|$ (EP wave vector). f) Real and imaginary parts of the mode energies (dots with error bars -- experiment, lines -- theory). \label{fig1}}
\end{center}
\end{figure}

The sample we study is an organic microcavity with metallic mirrors, shown in Fig.~\ref{fig1}(a)\cite{Ren20020}. The active layer is a microcrystal of a an organic molecule, 4,4'-bis[4-(di-p-tolylamino)styryl]biphenyl (DPAVBi), whose structure is shown in Fig.~\ref{fig1}(b). The properties of the sample are strongly anisotropic along the axes of the active microbelt crystal, also shown in the figure. 

We begin by showing the unpolarized reflectivity of the sample in the two orthogonal directions (Fig.~\ref{fig1}(c,d)). The reflectivity is plotted as a function of energy and wave vector $k_x$ and $k_y$. We focus on two particular eigenmodes, which exhibit the clearest behavior. First of all, we note that the two branches show very different effective masses and very different linewidths. This is due to the strongly polarized nature of excitons in DPAVBi (see Methods and Supplemental Figure 1, demonstrating the anisotropy of the excitonic absorption by the microbelt). The exciton ($E_x$~2.7 eV) strongly couples with the photonic modes only in the $H$ polarization (exhibiting a Rabi splitting of 80~meV), whereas the $V$-polarized modes remain unaffected by the excitonic resonance. The strongly-coupled modes exhibit a higher mass and a smaller linewidth, both because of their reduced photonic fraction.

In the two k-space directions, the behavior of the two modes is qualitatively different: a crossing of the weakly and strongly coupled polarization branches occurs in the $x$ direction and an anti-crossing in the $y$ direction. This anti-crossing is \emph{not} the result of the above mentioned strong exciton-photon coupling. It is rather due to the emergent optical activity of the structure, which becomes sufficiently large at the anticrossing wave vector. Optical activity has recently been shown to emerge at the macroscopic level in cavity structures, when the linear birefringence is so high that oppositely-polarized modes of opposite parity become degenerate \cite{Rechcinska2019}. It is therefore a coupling which occurs between the photonic part of the modes. This is illustrated by Supplementary Figure 2, showing (with a thicker sample) that the anticrossing only appears for opposite parity branches. Because of the time-reversal symmetry of the optical activity, it must change sign with the wave vector in one direction and is therefore zero along the perpendicular direction, which is why no anticrossing is observed along $k_x$.

The two closest branches can be described by a $2\times 2$ effective non-Hermitian Hamiltonian describing two polarization subbands with two different effective masses stemming from a different coupling with the exciton. The non-Hermitian contribution has to be included because of the difference of the linewidths.
The Hamiltonian is written in the \emph{linear} polarization basis:
\begin{equation}
\label{Ham2x2}
{H_0} = \left( {\begin{array}{*{20}{c}}
{{\beta _0} + (\xi  - \beta ){k^2} + i\Gamma-i\Gamma_0 }&{V{k_x}}\\
{V{k_x}}&{ - {\beta _0} + (\xi  + \beta ){k^2} - i\Gamma-i\Gamma_0 }
\end{array}} \right)
\end{equation}
where $\beta_0$ represents the splitting of the two modes at $k=0$, $\xi  ={\hbar ^2}/2{m^*}$ with ${m^*} = \frac{{{m_H}{m_V}}}{{{m_H} + {m_V}}}$, $V$ represents the emergent optical activity along the $k_x$ axis, $\Gamma$ is the half-difference of the broadenings of the modes, and $\Gamma_0$ is the half-average of the broadenings. Finally, $\beta$ is the difference of the effective masses of the two modes, which comes from the fact that one mode is coupled with the exciton, while there is no coupling for the other mode. The theoretical dispersions calculated with the Hamiltonian~\eqref{Ham2x2} are shown in Fig.~\ref{fig1}(c,d) with dashed lines. The best fit is obtained with the following parameters: $\beta_0=130$~meV, $\Gamma=22$ meV, $\beta=1$~meV/$\mu$m$^{2}$ , $m^{*}=9.97\times 10^{-6}~m_e$ and $V=1.8\times 10^{-6}$ meV.


The Hamiltonian is symmetric versus $k_y$ and anti-symmetric versus $k_x$.
Since the branches are crossing along $k_x$ and anticrossing along $k_y$, there are necessarily 4 points at which the transition between the two regimes occurs. These are the famous exceptional points characteristic for  non-Hermitian systems.  The plot of experimentally measured reflectivity along a circle of constant $|k|$ passing through one of the exceptional points is shown in Fig.~\ref{fig1}(e). The extracted mode energies and linewidths are shown in Fig.~\ref{fig1}(f) with points, and the corresponding real and imaginary parts of the theoretical eigenenergies appear as solid lines. In systems with perfectly balanced gain and losses, the exceptional points correspond to the transition between the PT-symmetric regime with real eigenvalues and the PT-broken regime with imaginary eigenvalues \cite{Bender1998}. The same transition is still present in our case, in spite of the overall decay $\Gamma_0$ and the observed behavior of the modes confirms the presence of a second order exceptional point at $k^*$.

\begin{figure}[tbp]
\begin{center}
\includegraphics[width=14.5cm]{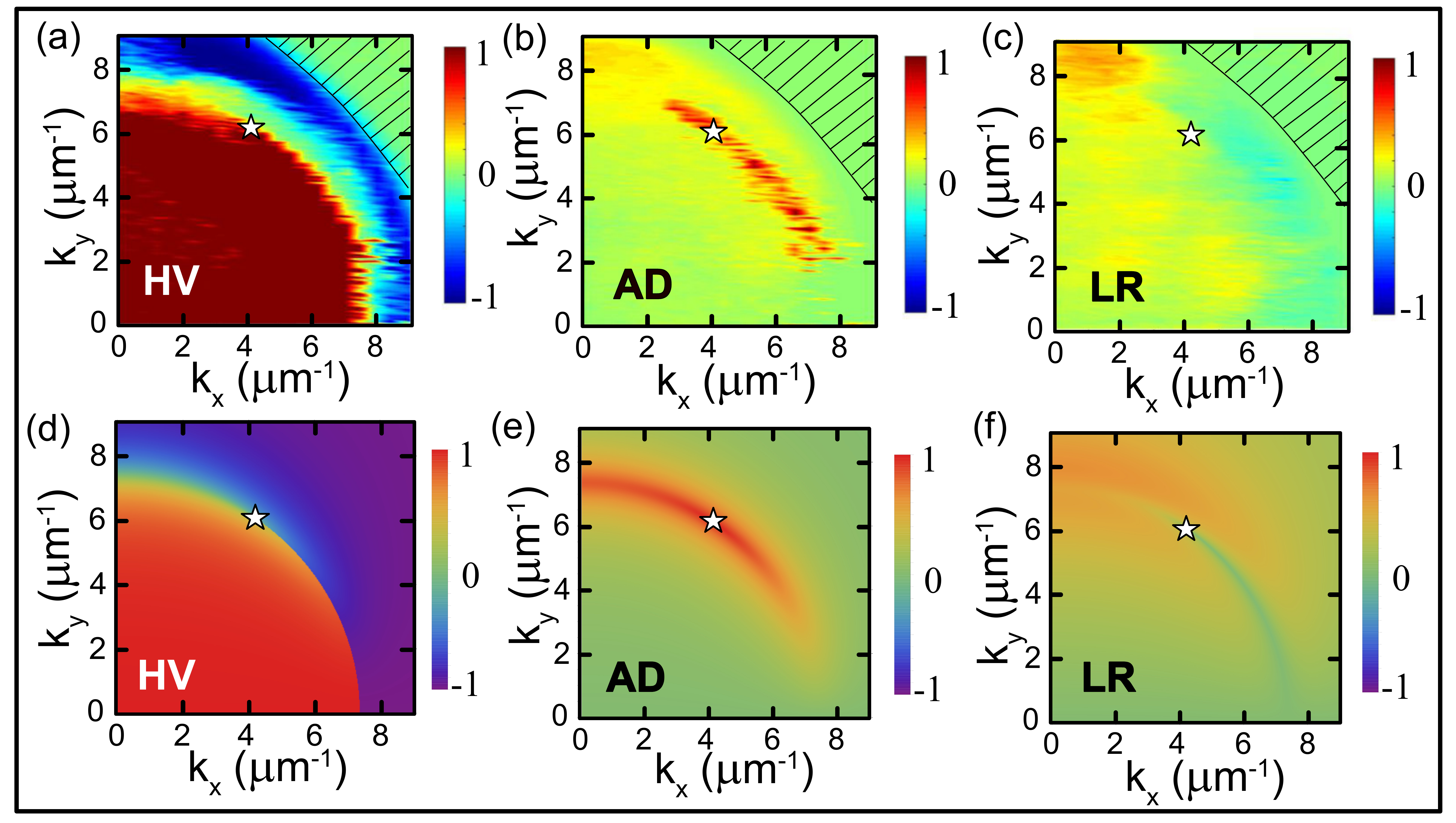}
\caption{\textbf{Stokes vector components of the lowest energy eigenstates (experiment and theory).} a)-c) experiment ($S_1$, $S_2$, $S_3$); d)-f) theory ($S_1$, $S_2$, $S_3$). A region where the pseudospin could not be extracted experimentally is hatched.  \label{fig2}}
\end{center}
\end{figure}

The eigenvalues do not tell everything about physical systems: the corresponding eigenstates are also important. While the famous Berry curvature and its integral, the Chern number, seem to be less relevant for non-Hermitian systems in the vicinity of exceptional points due to the essentially non-adiabatic behavior \cite{Berry2011a,Berry2011b}, other quantities linked with the eigenstates, such as the quantum metric, play a key role in the wavepacket or beam dynamics  \cite{Solnyshkov2020arxiv}.
The measurement of the Stokes vector for each eigenstate in reciprocal space \cite{Bleu2018effective,gianfrate2020measurement} allows to extract the quantum metric using the definition of the quantum geometric tensor (whose real part is the quantum metric, and the imaginary part is the Berry curvature):
\begin{equation}
    g_{ij}=\Re\left[\bra{\nabla\psi}\ket{\nabla\psi}-\bra{\psi}\ket{\nabla\psi}\bra{\nabla\psi}\ket{\psi}\right]
\end{equation}
where $\ket{\psi}$ is the eigenstate (written as a spinor, e.g. a Jones vector), and the gradient is taken in the parameter space (the reciprocal space). 
We now focus on a quarter of the reciprocal space containing a single exceptional point, and extract the Stokes vectors of the modes from polarization-resolved reflectivity measurements (See Methods). An energy spectrum is obtained in each of the six polarizations (H,V,D,A,L,R) for each point of the reciprocal space. We use a Lorentzian fit in order to get the positions, the relative intensities $I$, and the widths of the two modes, which permits the extraction of a 2D reciprocal space map of the Stokes vector components $S_1$, $S_2$, $S_3$ of the lower branch, shown in Fig.~\ref{fig2}(a-c). 
The validity of the effective $2\times 2$ Hamiltonian \eqref{Ham2x2} is confirmed by the good fit of the dispersions in Fig.~\ref{fig1}(c,d) and by the agreement between the experimentally extracted components of the Stokes vector (Fig.~\ref{fig2}(a-c)) and the theoretically calculated ones (Fig.~\ref{fig2}(d-f)). The EP located at $k_x^*=4.01~\mu $m$^{-1}$ and $k_y^*=6.12~\mu $m$^{-1}$ is shown by a white star. The two components $S_1$ and $S_3$ cancel at this point, while $S_2$ exhibits a maximum (similar to the circular polarization observed at the Voigt points).

\begin{figure}[tbp]
\begin{center}
\includegraphics[width=16cm]{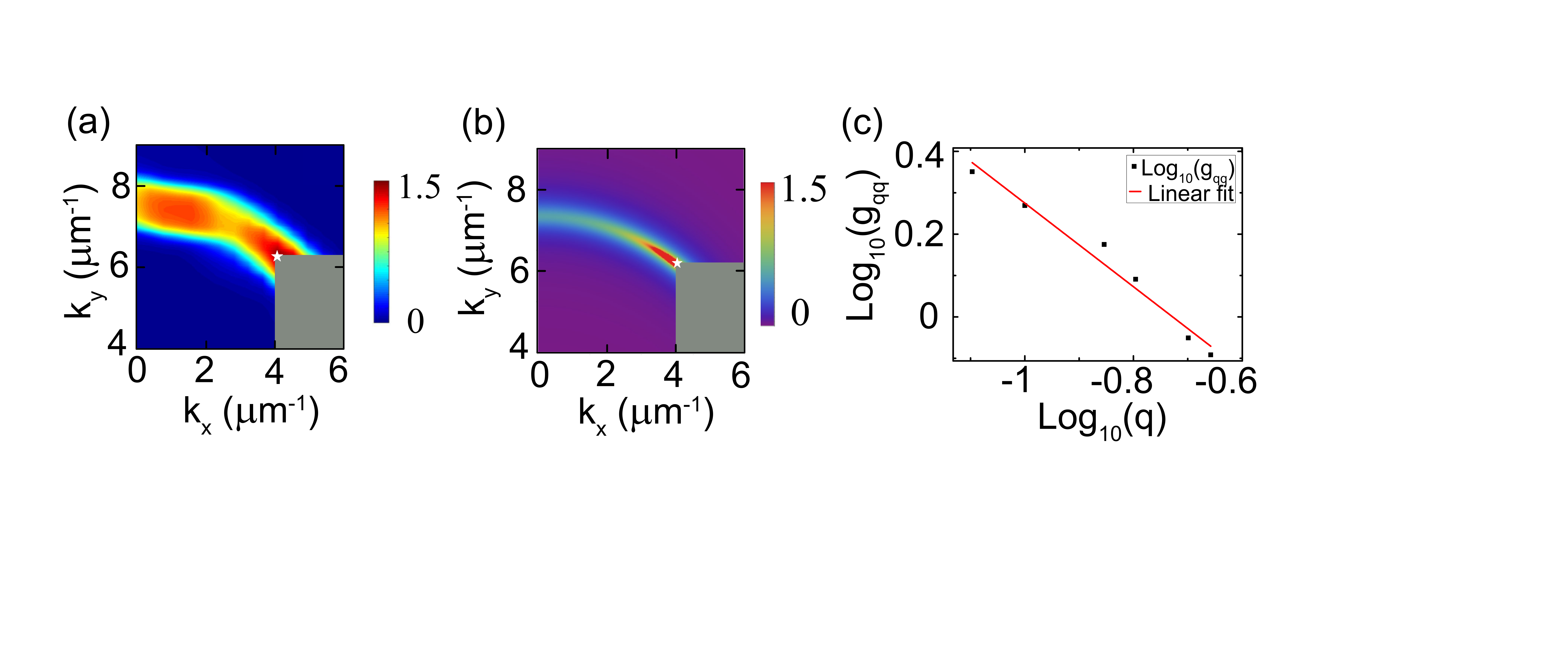}
\caption{\textbf{Quantum metric of an exceptional point.} 2D maps of the trace of the quantum metric in the vicinity of an EP: a) experiment; b) theory. The gray region covers the discontinuity of the wavefunction (branch cut). c) A log-log plot of the experimentally extracted quantum metric $q_{qq}$ and its fit, giving the scaling exponent $n=-1.01\pm0.08$. \label{fig3}}
\end{center}
\end{figure}

Once the Stokes vectors are known, one can extract the quantum metric elements as described in details in \cite{Bleu2018effective}.
The results of this extraction are shown as a 2D plot of the trace of the quantum metric $g_{xx}+g_{yy}$ in Fig.~\ref{fig3}(a). The part of the reciprocal space corresponding to the branch cut of the Riemann surface formed by the eigenstates is covered by a gray rectangle. A clear maximum is visible in the vicinity of the EP. The global behavior of the metric is in a good agreement with theoretical predictions based on the eigenstates of \eqref{Ham2x2}  (Fig.~\ref{fig3}b). 

The quantum metric is known to diverge hyperbolically at the exceptional points of the 2nd order (with 2 crossing branches) \cite{Brody2013, Solnyshkov2020arxiv}, and an explicit expression for the metric in the vicinity of the exceptional point can be written as:
\begin{equation}
   {g_{qq}} = \frac{{\sqrt {{{\alpha _x}^2{\cos }^2}\phi + {{\alpha _y}^2{\sin }^2}\phi } }}{{8\Gamma q}} + \frac{{{{\alpha _x}^2{\cos }^2}\phi  + {{\alpha _y}^2{\sin }^2}\phi }}{{16{\Gamma^2}}}
    \label{gqq}
\end{equation}
where $q$ is the wave vector measured from the exceptional point, and $\alpha_{x,y}$ is proportional to the difference of the group velocities at the crossing point (the celerity of the effective Dirac Hamiltonian, see Methods for details).
Experimentally, the values of the quantum metric are obtained only for a finite number of pixels in the reciprocal space, which can be close to the exceptional point, but never  fall on it exactly. In order to demonstrate the hyperbolic divergence, we choose a particular direction  in the reciprocal space, where the experimental resolution is the highest ($k_y$), and plot the quantum metric in log-log scale for several experimental points closest to the exceptional point as a function of $q=|k_y-k_y^*|$, which is the distance from this point. A fit with a power law $g_{qq}\sim q^n$ allows to determine the scaling of the quantum metric $n=-1.01\pm0.08$. We can therefore conclude that we have observed the hyperbolic divergence of the quantum metric of a 2nd-order exceptional point experimentally.

The agreement between the experiment and the theory can be checked further, by extracting the second (constant) term from the trace of the quantum metric and comparing it with the parameters of the effective Hamiltonian \eqref{Ham2x2}
obtained from the dispersions shown in Fig.~\ref{fig1}(c,d).
For this, we fit the experimentally extracted values of the quantum metric with a  function $f(q)$ corresponding to the reduced expression \eqref{gqq} of the quantum metric tensor $g_{qq}$:
\begin{equation}
    f=\frac{\eta}{q} + 4{\eta^2}
\end{equation} 
where $\eta  = \alpha_y/8\Gamma$. The fit gives us $\eta=0.173\pm0.004$ which can be compared to $\alpha$ and $\Gamma$ extracted by fitting the experimental dispersion. This fit gives $\Gamma= 4.57 \pm0.34$ meV and $\alpha_y  = 8 \pm2\ $meV/$\mu$m$^{-1}$, which gives $\eta_{exp}=0.21\pm 0.07\ \mu m^{-1} $. We can see that the value of $\eta$ obtained from fitting the extracted metric falls within the bounds of the confidence interval obtained from the analysis of the dispersion. This agreement validates both the metric extraction procedure and the theoretical analysis of the Hamiltonian and its eigenstates.

To conclude, we have studied exceptional points in an organic microcavity. We have extracted the Stokes vectors of the eigenstates in the vicinity of the exceptional point and then calculated the quantum metric tensor. Our measurements confirm that the quantum metric of a 2nd-order exceptional point exhibits a hyperbolic divergence. This is expected to affect the dynamics of wave packet (the trajectories of optical beams) at exceptional points.

\begin{methods}

\textbf{Microcavity fabrication}\\
The DPAVBi ($\>$98\%) was purchased from Aldrich and used directly without further purification. The DPAVBi microbelts were prepared by re-precipitation method, by injected 200 $\mu$L DPAVBi/THF into 2 mL Hexane under stirring. DPAVBi crystals can be obtained after standing for 0.5 h which dispersed in the Hexane solution. The microbelt’s width is around $20\mu$m, with the thickness of $2.0-3.0\mu$m and the length of around several hundreds of micrometers. Because of the crystal structure, the DPAVBi shows anisotropic excitonic absorption, which is demonstrated in Supplemental Figure 1.

To fabricate the microcavity, firstly, we vacuum thermally evaporate 80 ($\pm 5$) nm silver (Ag) (reflectivity: R $\geq 99\%$) on the glass substrate, the root mean square roughness (R$_{q}$) of the silver film in $5\mu$m$\times5\mu$m area is 2.4 nm, a 20 ($\pm$2) nm SiO$_{2}$ layer was deposited using vacuum Electron Beam evaporate on the silver film with R$_{q}$ of 2.31 nm. Then the prepared DPAVBi suspension were uniformly dispersed on the silver/SiO$_{2}$ film substrate, after the hexane evaporated, 20 ($\pm$2) nm SiO$_{2}$ layer and 35 ($\pm$2) nm (R $\approx$ 50\%) silver was fabricate to form the microcavity. The schematic of the DPAVBi microcavity is shown in Fig.~\ref{fig1}(a). The 20 nm SiO$_{2}$ layer is used to prevent the fluorescence quenching of the DPAVBi microbelt caused by directly contact of the metallic silver with the crystal\cite{Ren20020}. 

\textbf{Optical spectroscopy}\\
The angle-resolved spectroscopy was measured using a Halogen lamp with the 
wavelength range of 400-700 nm. The light source was entered and collected by 
using the $100\times$ microscope objective with a high aperture (0.95), the 
collection angle can achieve $\pm70^{\circ}$. The Momentum space of the 
reflectivity was located at the back focal plane of the objective lens. Lens 1-4 
formed a conjugate plane with the back focal plane of the objective lens, the k-
space light distribution can be imaged at the right focal plane of the L2 and L4 
on the spectrometer slit with a liquid-nitrogen-cooled CCD. Using four lenses here 
is to adjust the magnification of the final image and provide flexibility for 
effective light collection. By stepping moving the L4 lens, the image can be 
scanned on the slit for tomography to obtain spectrally resolved 2D k-space 
images. In order to investigate the polarization properties, we placed a linear 
polarizer, half-wave plate and a quarter-wave plate in front of the spectrometer 
slit to obtain the polarization state of each pixel in the k-space, horizontal-
vertical ($0^{\circ}$ and $90^{\circ}$),
 diagonal ($\pm45^{\circ}$)
  and circular 
($\sigma+$ and $\sigma-$) basis. The spectroscopy setup allows obtaining the 
polarization-resolved spectra of DPAVBi exciton absorption and polaritonic/cavity 
dispersion, as shown in Supplemental Figure 1 and 2.

\textbf{Theory}\\
We begin with a Hamiltonian describing two polarization subbands with two different effective masses stemming from a different coupling with the exciton. The Hamiltonian is written in the \emph{linear} polarization basis.
\[{H_0} = \left( {\begin{array}{*{20}{c}}
{{\beta _0} + (\xi  - \beta ){k^2} + i\Gamma }&{V{k_x}}\\
{V{k_x}}&{ - {\beta _0} + (\xi  + \beta ){k^2} - i\Gamma }
\end{array}} \right)\]
where $\beta_0$ represents the splitting of the two modes at $k=0$, $\xi  ={\hbar ^2}/2{m^*}$ with ${m^*} = \frac{{{m_H}{m_V}}}{{{m_H} + {m_V}}}$, $V$ represents the optical activity along the $k_x$ axis, and $\Gamma$ is the difference in the broadening of the modes. Finally $\beta$ is the difference of the effective masses of the two modes, which comes from the fact that one mode is coupled with the exciton, while there is no coupling for the other mode. Considering the exceptional point to be at $k_0=\{k_{x0},k_{y0}\}$, we apply series expansion to the Hamiltonian $H_0$ using $\overrightarrow k  = \overrightarrow {{k_0}}  + \overrightarrow {{q}}  $. The EP is located at the point, where the real part of the effective field is equal and perpendicular to the imaginary part of the effective field, which means here that $Vk_{x0}=\Gamma$, allowing to determine $k_{x0}$. The second condition reads $\beta k_0^2=\beta_0$. The resulting approximate Hamiltonian reads:

\[{H_1} = \left( {\begin{array}{*{20}{c}}
{\xi \frac{{{\beta _0}}}{\beta } + (\xi  - \beta )\left( {\frac{\Gamma }{V}{q_x} + \sqrt {\frac{{{\beta _0}}}{\beta } - \frac{{{\Gamma ^2}}}{{{V^2}}}} {q_y}} \right) + i\Gamma }&{\Gamma  + V{q_x}}\\
{\Gamma  + V{q_x}}&{\xi \frac{{{\beta _0}}}{\beta } + (\xi  + \beta )\left( {\frac{\Gamma }{V}{q_x} + \sqrt {\frac{{{\beta _0}}}{\beta } - \frac{{{\Gamma ^2}}}{{{V^2}}}} {q_y}} \right) - i\Gamma }
\end{array}} \right)\]
This Hamiltonian describes a tilted Dirac cone transformed into two exceptional points. To simplify it further, we remove the terms responsible for the tilt and keep only the terms relevant for the exceptional point: 

\[{H_2} = \left( {\begin{array}{*{20}{c}}
{-\beta \left( {\frac{\Gamma }{V}{q_x} + \sqrt {\frac{{{\beta _0}}}{\beta } - \frac{{{\Gamma ^2}}}{{{V^2}}}} {q_y}} \right) + i\Gamma }&{\Gamma  + V{q_x}}\\
{\Gamma  + V{q_x}}&{\beta \left( {\frac{\Gamma }{V}{q_x} + \sqrt {\frac{{{\beta _0}}}{\beta } - \frac{{{\Gamma ^2}}}{{{V^2}}}} {q_y}} \right) - i\Gamma }
\end{array}} \right)\]
i.e.
\[{H_2} = \left( {\Gamma  + V{q_x}} \right) {{\sigma _x}}  + \left( {-\beta \left( {\frac{\Gamma }{V}{q_x} + \sqrt {\frac{{{\beta _0}}}{\beta } - \frac{{{\Gamma ^2}}}{{{V^2}}}} {q_y}} \right) + i\Gamma } \right) {{\sigma _z}} \]
Here, the term $\beta\Gamma q_x/V$ is responsible only for the rotation of the system of coordinates around the exceptional point, and we can safely neglect it  in order to simplify the picture.
Therefore, $H_2$ can take the following simple shape:
\[{H_3} = \left( {\Gamma  + \alpha_x {q_x}} \right){\sigma _x} + \left( {i\Gamma  - \alpha_y {q_y}} \right){\sigma _z}\]
 For small q, the eigenstates can be written as 
\begin{equation}
   \left| {{\psi _q}} \right\rangle  = {\left( {1 - q/4\Gamma\sqrt {{{\cos }^2}(\phi ){\alpha _x}^2 + {{\sin }^2}(\phi ){\alpha _y}^2} ,\,\sqrt {q/2\Gamma} \sqrt {\cos (\phi ){\alpha _x} - i\,\sin (\phi ){\alpha _y}} } \right)^T} 
\end{equation}
which gives rise to a divergent symmetric distribution of the quantum metric around the exceptional point:
\begin{equation}
   {g_{qq}} = \frac{{\sqrt {{{\cos }^2}(\phi ){\alpha _x}^2 + {{\sin }^2}(\phi ){\alpha _y}^2} }}{{8\Gamma q}} + \frac{{{{\cos }^2}(\phi ){\alpha _x}^2 + {{\sin }^2}(\phi ){\alpha _y}^2}}{{16{\Gamma^2}}}
\end{equation}
We therefore extract the quantum metric from the experimental data in vicinity of the exceptional point ($q=0$), along a particular line in the reciprocal space where the experimental resolution is the best.

\end{methods}

\bibliographystyle{naturemag}
\bibliography{biblio}

\begin{thebibliography}{10}
\expandafter\ifx\csname url\endcsname\relax
  \def\url#1{\texttt{#1}}\fi
\expandafter\ifx\csname urlprefix\endcsname\relax\def\urlprefix{URL }\fi
\providecommand{\bibinfo}[2]{#2}
\providecommand{\eprint}[2][]{\url{#2}}

\bibitem{provost1980riemannian}
\bibinfo{author}{Provost, J.} \& \bibinfo{author}{Vallee, G.}
\newblock \bibinfo{title}{Riemannian structure on manifolds of quantum states}.
\newblock \emph{\bibinfo{journal}{Communications in Mathematical Physics}}
  \textbf{\bibinfo{volume}{76}}, \bibinfo{pages}{289--301}
  (\bibinfo{year}{1980}).

\bibitem{berry1989quantum}
\bibinfo{author}{Berry, M.}
\newblock \bibinfo{title}{The quantum phase, five years after}.
\newblock In \emph{\bibinfo{booktitle}{Geometric phases in physics}},
  \bibinfo{pages}{7} (\bibinfo{publisher}{World Scientific, Singapore},
  \bibinfo{year}{1989}).

\bibitem{Berry2020}
\bibinfo{author}{Berry, M.~V.} \& \bibinfo{author}{Shukla, P.}
\newblock \bibinfo{title}{Quantum metric statistics for random-matrix
  families}.
\newblock \emph{\bibinfo{journal}{Journal of Physics A: Mathematical and
  Theoretical}} \textbf{\bibinfo{volume}{53}}, \bibinfo{pages}{275202}
  (\bibinfo{year}{2020}).

\bibitem{Yu2019}
\bibinfo{author}{Yu, M.} \emph{et~al.}
\newblock \bibinfo{title}{{Experimental measurement of the quantum geometric
  tensor using coupled qubits in diamond}}.
\newblock \emph{\bibinfo{journal}{National Science Review}}
  \textbf{\bibinfo{volume}{7}}, \bibinfo{pages}{254--260}
  (\bibinfo{year}{2019}).
\newblock
  \eprint{https://academic.oup.com/nsr/article-pdf/7/2/254/32921980/nwz193.pdf}.

\bibitem{gianfrate2020measurement}
\bibinfo{author}{Gianfrate, A.} \emph{et~al.}
\newblock \bibinfo{title}{{Measurement of the quantum geometric tensor and of
  the anomalous Hall drift}}.
\newblock \emph{\bibinfo{journal}{Nature}} \textbf{\bibinfo{volume}{578}},
  \bibinfo{pages}{381--385} (\bibinfo{year}{2020}).

\bibitem{peotta2015superfluidity}
\bibinfo{author}{Peotta, S.} \& \bibinfo{author}{T{\"o}rm{\"a}, P.}
\newblock \bibinfo{title}{Superfluidity in topologically nontrivial flat
  bands}.
\newblock \emph{\bibinfo{journal}{Nature communications}}
  \textbf{\bibinfo{volume}{6}}, \bibinfo{pages}{8944} (\bibinfo{year}{2015}).

\bibitem{Liang2017}
\bibinfo{author}{Liang, L.}, \bibinfo{author}{Peotta, S.},
  \bibinfo{author}{Harju, A.} \& \bibinfo{author}{T\"orm\"a, P.}
\newblock \bibinfo{title}{Wave-packet dynamics of bogoliubov quasiparticles:
  Quantum metric effects}.
\newblock \emph{\bibinfo{journal}{Phys. Rev. B}} \textbf{\bibinfo{volume}{96}},
  \bibinfo{pages}{064511} (\bibinfo{year}{2017}).

\bibitem{Gao2014}
\bibinfo{author}{Gao, Y.}, \bibinfo{author}{Yang, S.~A.} \&
  \bibinfo{author}{Niu, Q.}
\newblock \bibinfo{title}{Field induced positional shift of {B}loch electrons
  and its dynamical implications}.
\newblock \emph{\bibinfo{journal}{Phys. Rev. Lett.}}
  \textbf{\bibinfo{volume}{112}}, \bibinfo{pages}{166601}
  (\bibinfo{year}{2014}).

\bibitem{Piechon2016}
\bibinfo{author}{Pi\'echon, F.}, \bibinfo{author}{Raoux, A.},
  \bibinfo{author}{Fuchs, J.-N.} \& \bibinfo{author}{Montambaux, G.}
\newblock \bibinfo{title}{Geometric orbital susceptibility: Quantum metric
  without {B}erry curvature}.
\newblock \emph{\bibinfo{journal}{Phys. Rev. B}} \textbf{\bibinfo{volume}{94}},
  \bibinfo{pages}{134423} (\bibinfo{year}{2016}).

\bibitem{PhysRevLett.115.166802}
\bibinfo{author}{Srivastava, A.} \& \bibinfo{author}{Imamoglu, A.}
\newblock \bibinfo{title}{Signatures of {B}loch-band geometry on excitons:
  Nonhydrogenic spectra in transition-metal dichalcogenides}.
\newblock \emph{\bibinfo{journal}{Phys. Rev. Lett.}}
  \textbf{\bibinfo{volume}{115}}, \bibinfo{pages}{166802}
  (\bibinfo{year}{2015}).

\bibitem{Moiseyev2011}
\bibinfo{author}{Moiseyev, N.}
\newblock \emph{\bibinfo{title}{{Non-Hermitian quantum mechanics}}}
  (\bibinfo{publisher}{Cambridge University Press (Cambridge, UK)},
  \bibinfo{year}{2011}).

\bibitem{Konotop2016}
\bibinfo{author}{Konotop, V.~V.}, \bibinfo{author}{Yang, J.} \&
  \bibinfo{author}{Zezyulin, D.~A.}
\newblock \bibinfo{title}{Nonlinear waves in $\mathcal{PT}$-symmetric systems}.
\newblock \emph{\bibinfo{journal}{Rev. Mod. Phys.}}
  \textbf{\bibinfo{volume}{88}}, \bibinfo{pages}{035002}
  (\bibinfo{year}{2016}).

\bibitem{Ganainy2018}
\bibinfo{author}{El-Ganainy, R.} \emph{et~al.}
\newblock \bibinfo{title}{{Non-Hermitian physics and PT symmetry}}.
\newblock \emph{\bibinfo{journal}{Nature Physics}}
  \textbf{\bibinfo{volume}{14}}, \bibinfo{pages}{11--19}
  (\bibinfo{year}{2018}).

\bibitem{Ozdemir2019}
\bibinfo{author}{{\"O}zdemir, {\c{S}}.}, \bibinfo{author}{Rotter, S.},
  \bibinfo{author}{Nori, F.} \& \bibinfo{author}{Yang, L.}
\newblock \bibinfo{title}{Parity--time symmetry and exceptional points in
  photonics}.
\newblock \emph{\bibinfo{journal}{Nature materials}}
  \textbf{\bibinfo{volume}{18}}, \bibinfo{pages}{783--798}
  (\bibinfo{year}{2019}).

\bibitem{Brody2013}
\bibinfo{author}{Brody, D.~C.} \& \bibinfo{author}{Graefe, E.~M.}
\newblock \bibinfo{title}{{Information geometry of complex hamiltonians and
  exceptional points}}.
\newblock \emph{\bibinfo{journal}{Entropy}} \textbf{\bibinfo{volume}{15}},
  \bibinfo{pages}{3361--3378} (\bibinfo{year}{2013}).
\newblock \eprint{1307.4017}.

\bibitem{chen2017exceptional}
\bibinfo{author}{Chen, W.}, \bibinfo{author}{{\"O}zdemir, {\c{S}}.~K.},
  \bibinfo{author}{Zhao, G.}, \bibinfo{author}{Wiersig, J.} \&
  \bibinfo{author}{Yang, L.}
\newblock \bibinfo{title}{Exceptional points enhance sensing in an optical
  microcavity}.
\newblock \emph{\bibinfo{journal}{Nature}} \textbf{\bibinfo{volume}{548}},
  \bibinfo{pages}{192--196} (\bibinfo{year}{2017}).

\bibitem{Solnyshkov2020arxiv}
\bibinfo{author}{Solnyshkov, D.~D.} \emph{et~al.}
\newblock \bibinfo{title}{Quantum metric and wavepackets at exceptional points
  in non-hermitian systems}.
\newblock \emph{\bibinfo{journal}{arXiv:2009.06987}}  (\bibinfo{year}{2020}).

\bibitem{Milburn2015}
\bibinfo{author}{Milburn, T.~J.} \emph{et~al.}
\newblock \bibinfo{title}{General description of quasiadiabatic dynamical
  phenomena near exceptional points}.
\newblock \emph{\bibinfo{journal}{Phys. Rev. A}} \textbf{\bibinfo{volume}{92}},
  \bibinfo{pages}{052124} (\bibinfo{year}{2015}).

\bibitem{Doppler2016}
\bibinfo{author}{Doppler, J.} \emph{et~al.}
\newblock \bibinfo{title}{Dynamically encircling an exceptional point for
  asymmetric mode switching}.
\newblock \emph{\bibinfo{journal}{Nature}} \textbf{\bibinfo{volume}{537}},
  \bibinfo{pages}{76--79} (\bibinfo{year}{2016}).

\bibitem{Bliokh2017}
\bibinfo{author}{Leykam, D.}, \bibinfo{author}{Bliokh, K.~Y.},
  \bibinfo{author}{Huang, C.}, \bibinfo{author}{Chong, Y.~D.} \&
  \bibinfo{author}{Nori, F.}
\newblock \bibinfo{title}{{Edge Modes, Degeneracies, and Topological Numbers in
  Non-Hermitian Systems}}.
\newblock \emph{\bibinfo{journal}{Phys. Rev. Lett.}}
  \textbf{\bibinfo{volume}{118}}, \bibinfo{pages}{040401}
  (\bibinfo{year}{2017}).

\bibitem{Shen2018}
\bibinfo{author}{Shen, H.}, \bibinfo{author}{Zhen, B.} \& \bibinfo{author}{Fu,
  L.}
\newblock \bibinfo{title}{{Topological Band Theory for Non-Hermitian
  Hamiltonians}}.
\newblock \emph{\bibinfo{journal}{Phys. Rev. Lett.}}
  \textbf{\bibinfo{volume}{120}}, \bibinfo{pages}{146402}
  (\bibinfo{year}{2018}).

\bibitem{Berry2011a}
\bibinfo{author}{Berry, M.} \& \bibinfo{author}{Uzdin, R.}
\newblock \bibinfo{title}{Slow non-hermitian cycling: exact solutions and the
  stokes phenomenon}.
\newblock \emph{\bibinfo{journal}{Journal of Physics A: Mathematical and
  Theoretical}} \textbf{\bibinfo{volume}{44}}, \bibinfo{pages}{435303}
  (\bibinfo{year}{2011}).

\bibitem{Berry2011b}
\bibinfo{author}{Berry, M.}
\newblock \bibinfo{title}{Optical polarization evolution near a non-hermitian
  degeneracy}.
\newblock \emph{\bibinfo{journal}{Journal of Optics}}
  \textbf{\bibinfo{volume}{13}}, \bibinfo{pages}{115701}
  (\bibinfo{year}{2011}).

\bibitem{Berry2003}
\bibinfo{author}{Berry, M.~V.} \& \bibinfo{author}{Dennis, M.~R.}
\newblock \bibinfo{title}{The optical singularities of birefringent dichroic
  chiral crystals}.
\newblock \emph{\bibinfo{journal}{Proc. R. Soc. Lond. A}}
  \textbf{\bibinfo{volume}{459}}, \bibinfo{pages}{1261} (\bibinfo{year}{2003}).

\bibitem{Voigt1902}
\bibinfo{author}{Voigt, W.}
\newblock \bibinfo{title}{On the behaviour of pleochroitic crystals along
  directions in the neighbourhood of an optic axis}.
\newblock \emph{\bibinfo{journal}{Philosophical Magazine Series}}
  \textbf{\bibinfo{volume}{4}}, \bibinfo{pages}{90} (\bibinfo{year}{1902}).

\bibitem{Richter2019}
\bibinfo{author}{Richter, S.} \emph{et~al.}
\newblock \bibinfo{title}{Voigt exceptional points in an anisotropic zno-based
  planar microcavity: Square-root topology, polarization vortices, and
  circularity}.
\newblock \emph{\bibinfo{journal}{Phys. Rev. Lett.}}
  \textbf{\bibinfo{volume}{123}}, \bibinfo{pages}{227401}
  (\bibinfo{year}{2019}).

\bibitem{Microcavities}
\bibinfo{author}{Kavokin, A.}, \bibinfo{author}{Baumberg, J.~J.},
  \bibinfo{author}{Malpuech, G.} \& \bibinfo{author}{Laussy, F.~P.}
\newblock \emph{\bibinfo{title}{Microcavities}} (\bibinfo{publisher}{Oxford
  University Press}, \bibinfo{year}{2011}).

\bibitem{Ren20020}
\bibinfo{author}{Ren, J.} \emph{et~al.}
\newblock \bibinfo{title}{Efficient bosonic condensation of exciton polaritons
  in an h-aggregate organic single-crystal microcavity}.
\newblock \emph{\bibinfo{journal}{Nano Lett.}} \bibinfo{pages}{in press}
  (\bibinfo{year}{2020}).

\bibitem{Rechcinska2019}
\bibinfo{author}{Rechci{\'n}ska, K.} \emph{et~al.}
\newblock \bibinfo{title}{Engineering spin-orbit synthetic hamiltonians in
  liquid-crystal optical cavities}.
\newblock \emph{\bibinfo{journal}{Science}} \textbf{\bibinfo{volume}{366}},
  \bibinfo{pages}{727--730} (\bibinfo{year}{2019}).
\newblock
  \eprint{https://science.sciencemag.org/content/366/6466/727.full.pdf}.

\bibitem{Bender1998}
\bibinfo{author}{Bender, C.~M.} \& \bibinfo{author}{Boettcher, S.}
\newblock \bibinfo{title}{Real spectra in non-hermitian hamiltonians having pt
  symmetry}.
\newblock \emph{\bibinfo{journal}{Phys. Rev. Lett.}}
  \textbf{\bibinfo{volume}{80}}, \bibinfo{pages}{5243--5246}
  (\bibinfo{year}{1998}).

\bibitem{Bleu2018effective}
\bibinfo{author}{Bleu, O.}, \bibinfo{author}{Malpuech, G.},
  \bibinfo{author}{Gao, Y.} \& \bibinfo{author}{Solnyshkov, D.~D.}
\newblock \bibinfo{title}{Effective theory of nonadiabatic quantum evolution
  based on the quantum geometric tensor}.
\newblock \emph{\bibinfo{journal}{Phys. Rev. Lett.}}
  \textbf{\bibinfo{volume}{121}}, \bibinfo{pages}{020401}
  (\bibinfo{year}{2018}).

\end{thebibliography}

\begin{addendum}
\item This work was supported by the National Key RD Program of China (Grant No. 2017YFA0204503 and 2018YFA0704805), the National Natural Science Foundation of China (12074303, 11804267, 21673144, 21790364, 21873065 and 21833005), the Beijing Natural Science Foundation of China (2192011), the High-level Teachers in Beijing Municipal Universities in the Period of 13th Five-year Plan (IDHT20180517 and CITTCD20180331), Beijing Talents Project (2019A23), the Open Fund of the State Key Laboratory of Integrated Optoelectronics (IOSKL2019KF01), Capacity Building for Sci-Tech Innovation-Fundamental Scientific Research Funds, Beijing Advanced Innovation Center for Imaging Theory and Technology.  
We acknowledge the support of the project "Quantum Fluids of Light"  (ANR-16-CE30-0021), of the ANR Labex Ganex (ANR-11-LABX-0014), and of the ANR program "Investissements d'Avenir" through the IDEX-ISITE initiative 16-IDEX-0001 (CAP 20-25). 
\item[Author contributions] J.~Ren -- investigation, formal analysis, visualization, methodology, writing; Q.~Liao -- conceptualization, funding acquisition, methodology, resources, supervision; F.~Li -- conceptualization, funding acquisition, methodology, supervision, writing, project administration; Y.~Li --methodology, writing; H.~Fu -- conceptualization, funding acquisition, methodology, resources, supervision; J.~Yao -- funding acquisition, supervision. D.~Solnyshkov -- conceptualization, funding acquisition, formal analysis, methodology, visualization, writing; G.~Malpuech -- conceptualization, funding acquisition, methodology, writing, supervision; C.~Leblanc -- formal analysis, conceptualization, validation, methodology, visualization, writing.
 \item[Competing Interests] The authors declare that they have no
competing financial interests.
 \item[Correspondence] Correspondence
should be addressed to liaoqing@cnu.edu.cn (Q.L); felix831204@xjtu.edu.cn (F.L.); dmitry.solnyshkov@uca.fr (D.S.); hbfu@cnu.edu.cn (H.F.)
\end{addendum}
\renewcommand{\thefigure}{S\arabic{figure}}

\begin{figure}[h]
\includegraphics[width=1\linewidth]{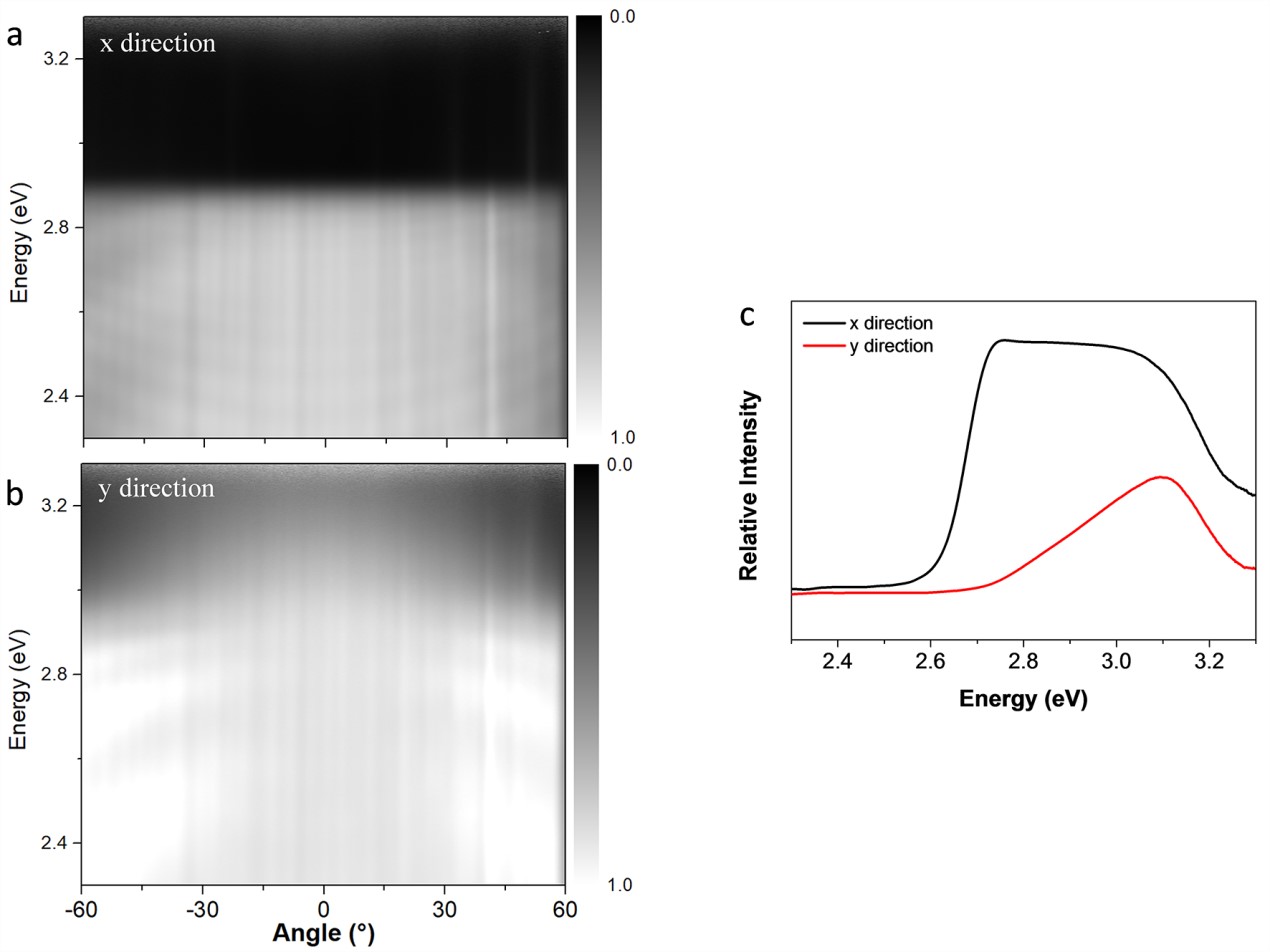}
\caption{\textbf{Anisotropy of the exciton absorption in DPAVBi}. Measured transmission of a bare DPAVBi crystal versus energy and wave vector along (a) X direction and (b) Y direction (as shown in Figure 2b), respectively. (c) A broad absorption peak with a maximum at $\sim$ 2.7 eV is observed (black line) when the polarization of the white light from a Halogen lamp is adjusted to be parallel to the belt length direction (X direction), while the absorption is much weaker (red line) when the polarization of the white light is vertical to the belt length direction (Y direction).}
\end{figure}

\begin{figure}[tbp]
\includegraphics[width=1\linewidth]{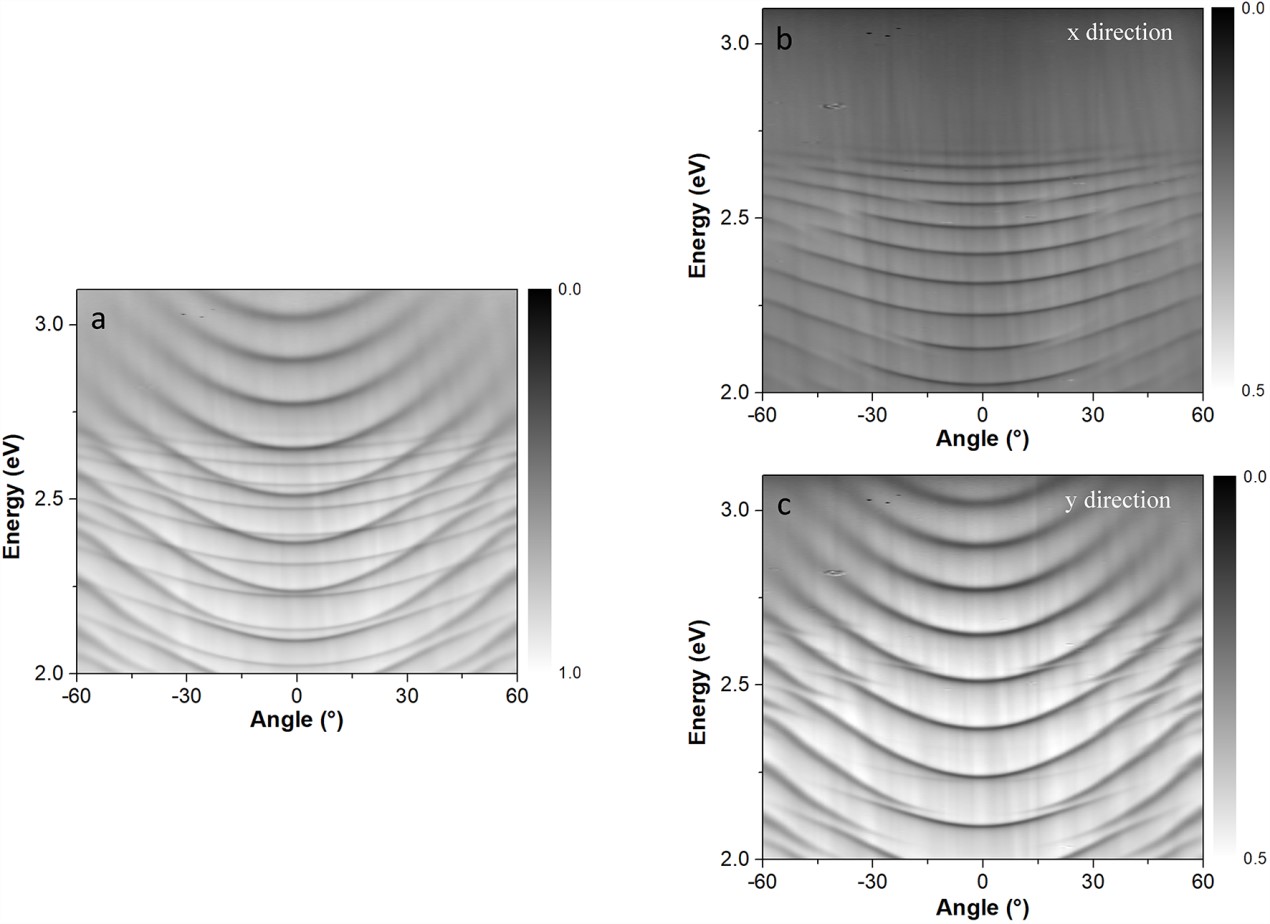}
\caption{\textbf{Branch anticrossings.}  Anticrossing only appears for branches of opposite parity. (a) Measured angle-resolved reflectivity spectrum of a microcavity with DPAVBi. Angle-resolved reflectivity detected along (b) X direction and (c) Y direction, showing a series of cavity modes with different dispersions, respectively. These distinct anisotropy features of cavity modes are consistent with the fact of the highly ordered uniaxial alignment of DPAVBi molecules in single-crystalline microbelts. Anticrossings of the modes of opposite parity and crossings of the modes of same parity are both observed in the experiment. }
\end{figure}

\begin{figure}[tbp]
\includegraphics[width=1\linewidth]{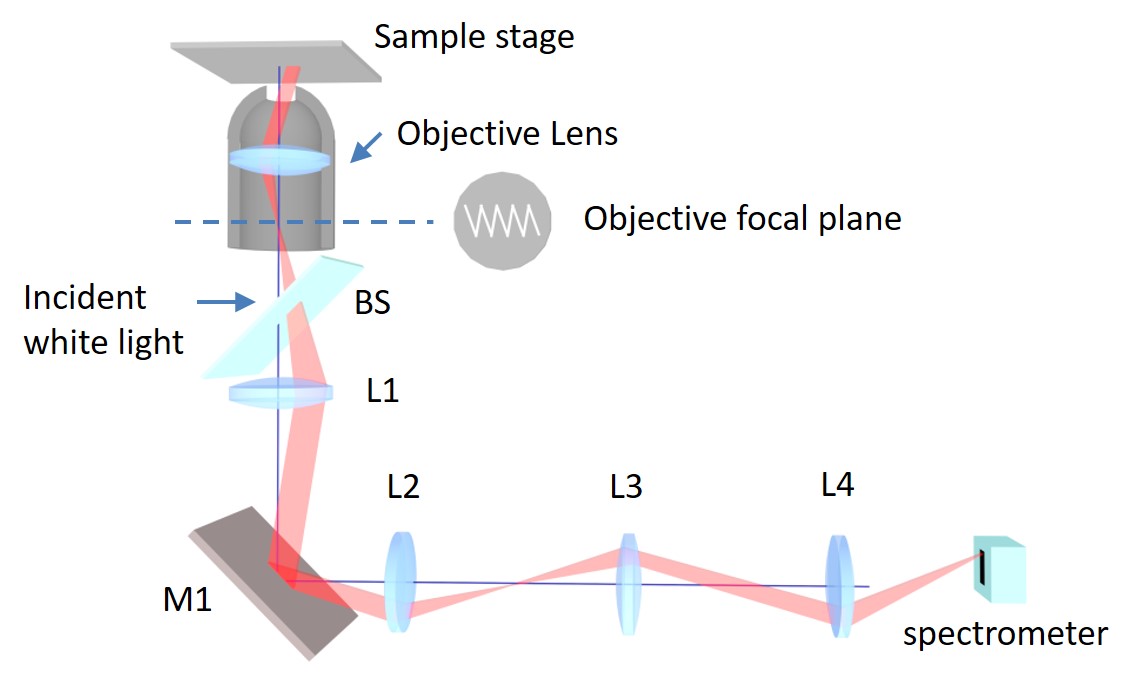}
\caption{\textbf{Schematic of the optical setup,} which has been explained in detail in Methods. }
\end{figure}

\end{document}